\documentclass[slac_one]{revtex4}
\usepackage{graphicx}
\usepackage{fancyhdr}
\begin{document}

\title{{\small{2005 ALCPG \& ILC Workshops - Snowmass,
U.S.A.}}\\ 
\vspace{12pt}
Potential for Higgs Physics at the LHC and Super-LHC} 

%

\author{K. S. Cranmer}
\affiliation{BNL, Upton, NY 11973, USA}

\begin{abstract}
The expected sensitivity of the LHC experiments to the discovery of
the Higgs boson and the measurement of its properties is presented in
the context of both the standard model and the its minimal
supersymmetric extension.  Prospects for a luminosity-upgraded
``Super-LHC'' are also presented.
\end{abstract}

\maketitle

\thispagestyle{fancy}


\section{Introduction}

The Large Hadron Collider (LHC) at CERN and the two multipurpose
detectors, {\sc Atlas} and {\sc CMS}, have been built in order to
discover the Higgs boson, if it exists, and explore the theoretical
landscape beyond the standard model~\cite{LHCC99-15,LHCC94-28}.  The
LHC will collide protons with unprecedented center-of-mass energy
($\sqrt{s}=14$ TeV) and luminosity ($10^{34}$ cm$^{-2}$s$^{-1}$); the
{\sc Atlas} and {\sc CMS} detectors will record these interactions
with $\sim \!10^8$ individual electronic readouts per event.  The first
collisions are expected in 2007, with a possible luminosity upgrade
around 2015.

Observation of the Higgs boson is key to confirming the description of
electroweak symmetry breaking in the standard model.  The standard
model Higgs sector has only one free parameter: the mass of the Higgs
boson, $m_H$.  Masses below 114.4 GeV/$c^2$ have been directly
excluded by LEP Higgs searches at the 95\%
confidence-level~\cite{Barate:2003sz}.  Indirect evidence of the Higgs
mass, through electroweak precision measurements, indicate a light
Higgs ($m_H \lesssim 185$ GeV/$c^2$), though the theory remains valid
until about 1~TeV$/c^2$ ~\cite{LEP-EW-WG2005-01}.

For a variety of reasons, it is reasonable to expect that
supersymmetry is manifest in nature.  The minimal supersymmetric
extension of the standard model (MSSM) requires an extended Higgs
sector with two Higgs doublets, corresponding to five physically
observable Higgs boson resonances.  The MSSM Higgs sector is typically
parametrized by the ratio of the vacuum expectation values of the two
doublets, $\tan \beta$, and the mass of the neutral, CP-odd Higgs
boson, $m_A$.  Large radiative corrections extend the upper-bound on
the mass of the lightest Higgs from its Born-level value $m_Z$ to
about 133 GeV/$c^2$.  Explicit CP-violation in the MSSM complicates
matters slightly~\cite{Carena:2000yi}.

At the LHC, production cross-section for the standard model Higgs
bosons falls from about 30 pb near the LEP limit to less than 0.1 pb
near $m_H \approx 1$ TeV.  The production is dominated by the
gluon-fusion process, followed by weak boson fusion, and associated
production with weak bosons and heavy quarks.  The decay is dominated
by heavy lepton pairs ($b\bar{b}$ and $\tau^+ \tau^-$), for masses
less than $2 m_W$, and pairs of weak bosons once above
threshold~\cite{Assamagan:2004mu}.  Due to the enormous rate of QCD
processes at the LHC, at least one high-$p_T$ lepton or photon or very
large missing $p_T$ is needed to trigger the event.\footnote{The
trigger requirement rules out the inclusive observation of $H\to
b\bar{b}$, {\it i.e.} from the dominant gluon-fusion production
process.}  Furthermore, due to the high design luminosity of the LHC,
an average of 23 soft p-p interactions are expected per bunch crossing
-- a phenomenon referred to as ``pile-up'' -- which makes the LHC
environment particularly challenging.  This pile-up effect will be
even more severe at the Super-LHC.

\section{LHC Discovery and Measurement Potential}

The Higgs discovery potential was detailed several years ago by both
{\sc Atlas} and {\sc CMS} assuming a total integrated luminosity of
300 fb$^{-1}$ per experiment~\cite{LHCC99-15,LHCC94-28}.  Both
experiments concluded that they should discover the standard model
Higgs with high significance for masses up to 1 TeV and measure the
mass of the Higgs boson to within 1\% across the entire mass range
(within 0.1\% for $m_H < 400$ GeV/$c^2$).  For $m_H > 200$ GeV/$c^2$,
non-standard spin and CP properties of the Higgs can be highly
excluded with 100 fb$^{-1}$~\cite{Buszello:2002uu}.  Observation of
two or more Higgs bosons would rule out a standard model Higgs sector,
and such observations are expected for many, though not all, regions
of the MSSM parameter space.\footnote{CP-violating scenarios have
regions of their parameter space in which the current set of search
channels are not sensitive.}

After the original assessments by {\sc Atlas} and {\sc CMS},
phenomenological studies indicated that weak boson fusion Higgs
production with decays to $W^+W^-$, $\tau^+\tau^-$, and $\gamma\gamma$
showed great potential for a discovery of the standard model
Higgs~\cite{Kauer:2000hi,Rainwater:1999sd,Plehn:1999xi}.  Within the
context of the CP-conserving MSSM, the complementary couplings to the
light and heavy CP-even neutral Higgs bosons allow the $\tau^+\tau^-$
channel to cover the entire $m_A$ -- $\tan\beta$
plane~\cite{Plehn:1999xi}.  These weak boson fusion analyses have now
been studied by the collaborations and provide for a more robust
discovery potential and improved coupling
measurements~\cite{Asai:2004ws,Abdullin:2005yn}.

Prior to the addition of weak boson fusion, the two most powerful
analyses for low-mass Higgs came from $H\to \gamma\gamma$ and $ttH\to
ttb\bar{b}$.  Since their initial study, great advances have been made
in terms of the Monte Carlo used to generate the particle-level
predictions.  More detailed study of the $ttH\to ttb\bar{b}$ channel
by {\sc Atlas} incorporating the systematic uncertainty on the
$b\bar{b}$ invariant mass spectrum indicate that the channel is not as
powerful as originally anticipated%
\footnote{It is not clear if
the {\sc Atlas} result is in agreement with the one from {\sc
cms}~\cite{CMS-2001-054}.}
 and is not sufficient for discovery~\cite{ATL-PHYS-2003-024}.  This
loss of sensitivity is found despite the use of multivariate
techniques.




By comparing the rates of different Higgs decays, it is possible to
measure various properties of the Higgs and how it couples to fermions
and bosons.  The interpretation of these measurements is tightly
coupled to the theoretical assumptions one makes.  Ratios of partial
widths and (with mild theoretical assumptions) absolute couplings can
typically be measured to an accuracy of
10-40\%~\cite{Duhrssen:2004cv}.  Some improvements to this result are
foreseen by reducing the uncertainty on the rate of $gg\to Hgg$ by
taking advantage of its characteristic $\phi_{jj}$ shape
\cite{DelDuca:2001fn} and arguing that certain theoretical
uncertainties may largely cancel in ratio~\cite{Anastasiou:2005pd}.

\section{Higgs Prospects at the Super-LHC}

While the prospects for the LHC are quite encouraging, there are many
measurements that are statistics-limited.  For example, observation of
$H\to \mu \mu$ (a non-third-generation fermion coupling) is straight
forward experimentally, but limited by the small branching
ratio~\cite{Plehn:2001qg,Han:2002gp}.  Similarly, additional
luminosity would increase the supersymmetric parameter space in which
two or more Higgs bosons are visible.  Also of great importance is the
measurement of the Higgs self-couplings, which is largely limited by
the integrated luminosity.

For the reasons stated above, a luminosity upgrade to the LHC is being
considered.  The so-called ``Super-LHC'', or SLHC, would have a design
luminosity of $10^{35}$ cm$^{-2}$s$^{-1}$ and would potentially gather
3000 fb$^{-1}$ of data~\cite{Bruning:2002yh}.  While details of such
an upgrade remain uncertain, it would most likely start around 2015.
Significant studies are already underway to assess the requisite
modifications to the accelerator, detector subsystems, and trigger and
data acquisition infrastructure.  Studies of the physics potential of
the SLHC by the experimental collaborations are somewhat limited, but
they do show the expected gains in coupling measurements and extended
discovery potential in the $m_A$ -- $\tan\beta$
plane~\cite{Gianotti:2002xx}.  Additionally, it has been shown in
phenomenological studies that the tri-linear Higgs self-coupling could
be measured to 20-30\%, and both experiments are working to confirm
this result~\cite{Baur:2002rb}. Unfortunately, small variations in the
tri-linear coupling will swamp variations in the quartic-coupling,
leaving little hope to measure the quartic coupling at any foreseeable
hadron collider~\cite{Plehn:2005nk}.

The existing studies show an enticing physics program, but with
attention turning to the start-up operations of the LHC, additional
studies are not likely to be forthcoming.  After some experience with
pile-up at the LHC design luminosity and potential observations of new
physics, the LHC collaborations will undoubtedly embark on more
studies germane to the SLHC physics potential.

\section{Conclusion}

If it exists, the LHC should discover standard model Higgs boson,
measure its mass accurately, and make various measurements of its
couplings, spin and CP properties.  In the context of the
CP-conserving MSSM, the LHC should be able to discover one or more
Higgs bosons over the entire $m_A$ -- $\tan\beta$ plane, with two or
more observable in many cases.  The large number of channels available
insure a robust discovery and offer many opportunities for additional
measurements.

Observation of $H\to \mu \mu$, measurement of the tri-linear Higgs
self-coupling, and various search channels are statistics-limited, and
only possible with a luminosity upgrade.  A luminosity upgrade would
substantially improve some of the coupling measurements and generally
extend the sensitivity in the MSSM Higgs plane.  Efforts are ongoing
to understand the upgrade of the LHC to the Super-LHC.

\section*{Acknowledgments}

I would like to thank David Rainwater and Tilman Plehn for the useful
discussions, Wesley Smith and Michael D\"uhrssen for their
contributions, and the organizers of Snowmass for their very enjoyable
conference.

This manuscript has been authored by Brookhaven Science Associates,
LLC under Contract No. DE-AC02-98CH1-886 with the U.S. Department of
Energy.  The U.S. Government retains, and the publisher, by accepting
the article for publication, acknowledges, a world-wide license to
publish or reproduce the published form of this manuscript, or allow
others to do so, for the U.S. Government purposes.





\end{document}